\documentclass[preprint,12pt]{elsarticle}
\usepackage{amsmath}
\usepackage{amssymb}
\usepackage{tikz}
\usetikzlibrary{decorations.markings}

\journal{Physics Letters B}

\begin{document}

\begin{frontmatter}



\title{On the cancellation of Newtonian singularities in higher-derivative gravity}


\author{Breno L. Giacchini}
\ead{breno@cbpf.br}

\address{Centro Brasileiro de Pesquisas F\'{\i}sicas -- CBPF,
Rua Dr. Xavier Sigaud 150, Urca, 22290-180, Rio de Janeiro, RJ, Brazil}

\begin{abstract}
Recently there has been a growing interest in quantum gravity theories with more than four derivatives, including both their quantum and classical aspects. In this work we extend the recent results concerning the non-singularity of the modified Newtonian potential to the most relevant case in which the propagator has complex poles. The model we consider is Einstein-Hilbert action augmented by curvature-squared higher-derivative terms which contain polynomials on the d'Alembert operator. We show that the classical potential of these theories is a real quantity and it is regular at the origin despite the (complex or real) nature or the multiplicity of the massive poles. The expression for the potential is explicitly derived for some interesting particular cases. Finally, the issue of the mechanism behind the cancellation of the singularity is discussed; specifically we argue that the regularity of the potential can hold even if the number of massive tensor modes and scalar ones is not the same.
\end{abstract}


\end{frontmatter}


\section{Introduction}
\label{intro}

The twentieth century has brought two fundamental ideas to Physics: the curved space-time and the quantization of matter. In spite of the great success each of these insights has achieved owed to the outstanding experimental verification of general relativity and quantum field theory, no fully consistent way of combining both paradigmatic theories is known. Since the 1960's it is known that the renormalization of quantum fields on curved space-time using perturbative methods requires introducing the curvature-squared terms $R^2, R_{\mu\nu}^2$ and $R_{\mu\nu\alpha\beta}^2$ which violate the unitarity of the theory~\cite{UtDW}. The situation is no better when gravity itself is quantized --- for example, the gravity model with fourth-derivative terms is renormalizable, but has negative-norm states~\cite{Stelle77}. Conciliating unitarity and renormalizability is one of the main problems in quantum gravity and has motivated the search for theories which relied on fundamentally different basic principles, such as string theory.

Nonetheless, within the framework of standard quantum field theory, the introduction of terms of order two in curvature but with more than four derivatives has shown to make the theory superrenormalizable~\cite{AsoreyLopezShapiro97} and also allowed for the possibility of providing a unitary S-matrix~\cite{ModestoShapiro16}. In fact, in this case the associated propagator may admit massive complex poles; such virtual modes would have complex kinetic energy, being unstable and leading to a unitary theory \textit{a la} Lee-Wick~\cite{ModestoShapiro16,Modesto2016}.

Renormalizability in gravity might be related to the behaviour of the classical interparticle potential of the model~\cite{Newton-high}; indeed there is a conjecture which states that renormalizable gravity theories have a finite non-relativistic potential at the origin~\cite{Accioly13_riddle}. This relation was first noticed in Stelle's seminal works~\cite{Stelle77,Stelle78} which showed that the fourth-derivative gravity is renormalizable, and has a regular potential. More recently, there have been interesting investigations on this conjecture in massive gravity models, and also in theories with dimensions different than four (see~\cite{Accioly13_riddle} and references therein).

In a recent paper~\cite{Newton-high} Modesto, Paula Netto and Shapiro moved a step further on this discussion with the generalization of Stelle's result so as to account for a class of superrenormalizable particular cases of the model defined by the action\footnote{The cosmological constant is omitted because it does not affect the regularity of the classical potential, besides being very small. Of course, the corresponding term is necessary to the renormalization of the theory. Our sign convention is to define the Minkowski metric as $\eta_{\mu\nu} = \text{diag} (1,-1,-1,-1)$, while the Riemann and Ricci tensors are ${R^\rho}_{\lambda\mu\nu} = \partial_\mu {\Gamma^\rho}_{\lambda\nu} + \cdots$ and $R_{\mu\nu} = {R^\rho}_{\mu\nu\rho}$. To simplify notation we set $\hbar = c = 1$.}~\cite{AsoreyLopezShapiro97}
\begin{eqnarray} \label{N_OrderGravity}
\mathcal{S}_\text{grav} &=& \int d^4 x \sqrt{-g} \, \bigg( \, \dfrac{2}{\kappa^2} R + R \tilde{F_1}^{(a)}(\square) R 
\nonumber
\\
&& + \, R_{\mu\nu} \tilde{F_2}^{(b)}(\square) R^{\mu\nu} + R_{\mu\nu\alpha\beta} \tilde{F_3}^{(c)}(\square) R^{\mu\nu\alpha\beta}  \,\bigg) \, ,
\end{eqnarray}
where $\kappa^2 = 32 \pi G$  and $\tilde{F}_{i}^{(j)}(\square)$ are polynomials of degree $j \geq 0$ on the d'Alembert operator. We recall that the condition $a=b=c$ makes the model renormalizable\footnote{The case $a=b=c=0$ corresponds to Stelle's renormalizable model~\cite{Stelle77}.}, and that it becomes superrenormalizable if $a=b=c \geq 1$. The analysis in~\cite{Newton-high} was restricted to the (super)renormalizable case with the additional constraint that the polynomials $\tilde{F}_{i}^{(j)}$ yielded only simple, real poles in the propagator. The main result they obtained was another verification of the aforementioned conjecture, showing that this model has a finite classical potential at the origin: the massive tensor and scalar modes contribute in such a manner to precisely cancel out the Newtonian singularity. Also, it was suggested that this happens not only because of a particular balance between the attractive forces owed by the healthy modes and the repulsive ones related to the ghosts, but also due to a specific matching of the number of tensor and scalar modes. Namely, they conjectured that if the number of these massive excitations is not the same (which is related to having polynomials $\tilde{F}_{i}^{(j)}$ of different orders and, therefore, losing renormalizability), the potential would not be regular.

In the present work we extend the result of~\cite{Newton-high} to the most general and more interesting case in which the massive poles of the propagator are complex, and may have degeneracies. We show that the action~\eqref{N_OrderGravity} always yields a real, regular potential at the origin --- disregard the number, the nature (complex or real) and order of the massive poles. As a consequence, the mechanism which allows the cancellation of the Newtonian singularity is broader than the one proposed in~\cite{Newton-high}.% The expression for the potential is also obtained for some interesting particular cases which appear throughout the work.

To close this introductory section, it is worthwhile to mention that there exists a connection between the polynomial action~\eqref{N_OrderGravity} and the infinite-derivative ``ghost free'' gravity~\cite{Tseytlin95Tomboulis} (see, for instance, Refs.~\cite{Non-local} for recent studies on singularities in the latter model). In fact, it was argued in~\cite{CountingGhosts} that the quantum corrections to the non-local classical model would lead to an infinite amount of complex ghost-like states, making the study of the polynomial gravity mandatory --- with a special interest to the case of complex poles. In particular, it was conjectured that the cancellation of the Newtonian singularity in the non-local model could be owed to the effect of an infinite number of ``hidden'' complex excitations~\cite{CountingGhosts}; the present work can be viewed as a step towards this result. Finally, it is good to remember that singularities (of both black hole type and cosmological) constitute a central topic in gravitational physics and one of the main reasons for quantum gravity. Hence it is important to explore the classical singularities in the new promising quantum gravity model~\eqref{N_OrderGravity}. The influence of fourth derivatives has already been investigated in the context of black holes~\cite{BlackHoles4} and cosmological solutions~\cite{Anderson}. Inasmuch as the general polynomial theory is more complicated than Stelle's gravity, it is sound to start from the Newtonian case, as it is done in the present work. Other investigations on the low-energy phenomenology of (super)renormalizable higher-derivative local theories are carried out in~\cite{AHGH15} and in the parallel works~\cite{Large,Seesaw}. 

%To simplify notation we set $\hbar = c = 1$. Our Minkowski metric is $\eta_{\mu\nu} = \text{diag} (1,-1,-1,-1)$, while the Riemann and Ricci tensors are ${R^\rho}_{\lambda\mu\nu} = \partial_\mu {\Gamma^\rho}_{\lambda\nu} + \cdots$ and $R_{\mu\nu} = {R^\rho}_{\mu\nu\rho}$.

%Our sign convention is to define the Minkowski metric as $\eta_{\mu\nu} = \text{diag} (1,-1,-1,-1)$, while the Riemann and Ricci tensors are ${R^\rho}_{\lambda\mu\nu} = \partial_\mu {\Gamma^\rho}_{\lambda\nu} + \cdots$ and $R_{\mu\nu} = {R^\rho}_{\mu\nu\rho}$. To simplify notation we set $\hbar = c = 1$.

\section{Real potential with complex poles}
\label{sec2}

The classical potential of a gravitational theory is computed by considering the metric to be a small fluctuation around the flat space-time, $g_{\mu\nu} = \eta_{\mu\nu} + \kappa^2 h_{\mu\nu}$, and approximating the geometric quantities by their linearized forms. The quadratic terms in the Riemann tensor need not to be considered in the linear approximation, because the relation ($p \in \mathbb{N}$)
\begin{equation*}
\int d^4 x \sqrt{-g} \left( R \square^p R - 4 R_{\mu\nu} \square^p R^{\mu\nu} + R_{\mu\nu\alpha\beta} \square^p R^{\mu\nu\alpha\beta} \right) = \mathcal{O}(h^3) 
%\quad \forall p \in \mathbb{N}
\end{equation*}
means that at this level there are only two independent quantities among the scalars $R\square^p R$, $R_{\mu\nu} \square^p R^{\mu\nu}$ and $R_{\mu\nu\alpha\beta} \square^p R^{\mu\nu\alpha\beta}$ (see, e.g., ~\cite{AsoreyLopezShapiro97}). Hence, we may substitute the polynomials $\tilde{F}_{i}^{(j)}$ by $F_1^{(p)} \equiv \tilde{F}_1^{(a)} - \tilde{F}_3^{(c)}$, $F_2^{(q)} \equiv \tilde{F}_2^{(b)} + 4 \tilde{F}_3^{(c)}$ and $F_3 \equiv 0$, which simplifies the Lagrangian associated with the action~\eqref{N_OrderGravity} leading to
\begin{equation} \label{Lagrangian}
\mathcal{L}_\text{grav} = \sqrt{-g} \left( \dfrac{2}{\kappa^2} R + R F_1^{(p)}(\square) R + R_{\mu\nu} F_2^{(q)}(\square) R^{\mu\nu} \right) ,
\end{equation}
where $p = \max \lbrace a, c\rbrace$ and $q = \max \lbrace b, c \rbrace$.

We note that, via the substitution $\partial_\mu \longmapsto - i k_\mu$, each $F_i^{(j)}(\square)$ corresponds to a polynomial $F_i^{(j)}(-k^2)$ in the momentum space representation. Let us now define the polynomials\\
\begin{equation} \label{Polynomials}
\begin{split}
Q_0 (k^2)  = & \,\, 1 - \kappa^2 k^2  \left[ F_2^{(q)}(-k^2) + 3 F_1^{(p)}(-k^2) \right] \, , \\
Q_2 (k^2)  = & \,\, 1 + \frac{\kappa^2 k^2}{2} F_2^{(q)}(-k^2) \, ,
\end{split}
\end{equation} 
respectively of order $n_0 = 1 + \max\lbrace p, q \rbrace$ and $n_2 = q+1$ on $k^2$. It is not difficult to verify that in the de~Donder gauge the momentum space representation of the propagator associated to~\eqref{Lagrangian} is given in terms of $Q_0$ and $Q_2$ as
\begin{equation}\label{propagator}
\begin{split}
D &  =  \frac{1}{k^2 Q_2} P^{(2)} - \frac{1}{2k^2 Q_0 } P^{(0-s)} + \frac{2\lambda}{k^2} P^{(1)} 
\\
& + \,  \left[- \frac{3}{2k^2 Q_0 } + \frac{4\lambda}{k^2} \right] P^{(0-w)} 
  +  \dfrac{\sqrt{3}}{2k^2 Q_0 } \left[ P^{(0-sw)} + P^{(0-ws)} \right] .
\end{split}
\end{equation} 
Here $\lambda$ is a gauge parameter, and $P^{(2)}$, $P^{(0-s)}$, etc. are the usual Barnes-Rivers operators~\cite{book}, whose indices have been omitted. The masses of the propagated fields correspond to the poles of~\eqref{propagator}, which turn out to be the roots of $Q_{0,2}$. According to the fundamental theorem of algebra, there are $n_0$ massive modes of spin-0, and $n_2$ massive spin-2 modes (complex roots and degeneracies may occur depending on the coefficients of the polynomials). Therefore, it is more useful to rewrite these polynomials in the factored form% which shows explicitly the poles of the propagator:
\begin{eqnarray} \label{polynomial_factor}
Q_i(k^2) = \dfrac{(m_{(i)1}^2 - k^2) (m_{(i)2}^2 - k^2) \cdots (m_{(i)n_i}^2 - k^2)}{m_{(i)1}^2 m_{(i)2}^2\cdots m_{(i)n_i}^2}.
\end{eqnarray}
The poles of the propagator~\eqref{propagator} are defined as $m_{(i)j}^2 \,$, or $\pm m_{(i)j}$ if we consider that the polynomial is on $k$. The index $i=0,2$ between parentheses labels the spin of the particle associated to the $j$\textsuperscript{th} excitation.

The field $h_{\mu\nu}$ generated by a point-like mass $M$ in rest, $T_{\mu\nu}(\textbf{r}) = M \eta_{\mu 0} \eta_{\nu 0} \delta^{(3)} (\textbf{r})$, can be evaluated by means of the Fourier transform method~\cite{Modesto2016,Newton-high} or via an auxiliary field formulation as in~\cite{Large}. The classical potential $\phi$ is then proportional to $h_{00}$, namely, $\phi = \frac{\kappa^2}{2} h_{00}$. It can also be computed following the scheme of~\cite{AcciolyPotential}. All in all, it is possible to show that the (modified) Newtonian potential is given by
\begin{eqnarray}
\phi(r) = -\frac{iGM}{\pi r} \int_{-\infty}^{+\infty} \frac{dk}{k} e^{ikr} \left( \dfrac{4}{3 Q_2(-k^2)} - \frac{1}{3 Q_0(-k^2)} \right) . \label{phi}
\end{eqnarray}

\begin{figure}[t!] 
\centering
 \begin{tikzpicture}[scale=2.4]
 \draw[->] (-1.2,0) -- (1.2,0) coordinate (x axis) node[above]{Re $z$};
 \draw[->] (0,-1.1) -- (0,1.2) coordinate (y axis) node[right]{Im $z$};;
 \draw[thick,black,decoration={markings, mark=at position 0.25 with {\arrow{>}}}, postaction={decorate}] (1,0) arc (0:180:1);
 \draw[thick,black] (0.5mm,0mm) arc (0:180:0.5mm);
 \draw[thick,black] (-1,0.0) -- (-0.5mm,0.0);
 \draw[thick,black] (0.5mm,0) -- (1,0);
 \node at (0.8,0.8) (nodeA) {$\Gamma$};

%PONTOS
 \draw (0,0) node[circle,fill,inner sep=1pt](a){};
 
 \draw (0,0.3) node[circle,fill,inner sep=1pt](a){};
 \draw (0,0.63) node[circle,fill,inner sep=1pt](a){};
 \draw (0,0.9) node[circle,fill,inner sep=1pt](a){};
 \draw (0,-0.3) node[circle,fill,inner sep=1pt](a){};
 \draw (0,-0.63) node[circle,fill,inner sep=1pt](a){};
 \draw (0,-0.9) node[circle,fill,inner sep=1pt](a){};
 
 \draw (0.1,0.7) node[circle,fill,inner sep=1pt](a){};
 \draw (0.5,0.2) node[circle,fill,inner sep=1pt](a){};
 \draw (0.75,0.3) node[circle,fill,inner sep=1pt](a){};
 \draw (0.1,-0.7) node[circle,fill,inner sep=1pt](a){};
 \draw (0.5,-0.2) node[circle,fill,inner sep=1pt](a){};
 \draw (0.75,-0.3) node[circle,fill,inner sep=1pt](a){};
 
 \draw (-0.1,0.7) node[circle,fill,inner sep=1pt](a){};
 \draw (-0.5,0.2) node[circle,fill,inner sep=1pt](a){};
 \draw (-0.75,0.3) node[circle,fill,inner sep=1pt](a){};
 \draw (-0.1,-0.7) node[circle,fill,inner sep=1pt](a){};
 \draw (-0.5,-0.2) node[circle,fill,inner sep=1pt](a){};
 \draw (-0.75,-0.3) node[circle,fill,inner sep=1pt](a){};
 \end{tikzpicture}
\caption{The contour $\Gamma$ used to evaluate the integral~\eqref{Integral2}. The poles on the upper half-plane occur at $z = + i m_{(i)j}$; while those on the lower one are located at $z = - i m_{(i)j}$. If $m_{(i)j}$ is a real root of the polynomial $Q_i$, the corresponding pole lies over the imaginary axis. The pairs of symmetric poles with respect to the imaginary axis represent the complex conjugate roots $m_{(i)j}$ and $\overline{m}_{(i)j}$. We do not consider massive poles for $z$ over the real axis since they are associated to tachyonic modes.}  \label{figure1}
 \end{figure}
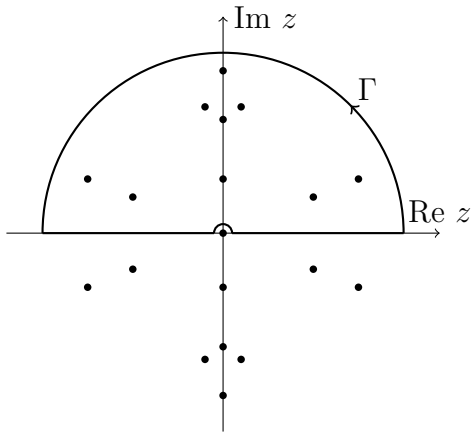

Accordingly, the fundamental integral to be evaluated is
\begin{equation} \label{Integral}
I_i = \int_{-\infty}^{+\infty} \frac{dk}{kQ_i(-k^2)} e^{ikr} ,
\end{equation}
which can be done using the contour technique and Cauchy's residue theorem, as in~\cite{Newton-high}. The assumption used in this reference was that the poles of the propagator were real and simple; in what follows we show that the same procedure can be generalized and applied if the poles are complex and of arbitrary order. To evaluate~\eqref{Integral} we proceed the analytic continuation of $k$ into the complex plane and write
\begin{equation} \label{Integral2}
I_i = 
%\int_{-\infty}^{+\infty} \frac{dk}{kQ_i(-k^2)} e^{ikr} = 
\int_\Gamma \frac{dz}{z Q_i(-z^2)} e^{izr} = \int_\Gamma dz \frac{e^{izr}}{z} \prod_{j=1}^{n_i} \frac{m_{(i)j}^2}{z^2 + m_{(i)j}^2},
\end{equation}
where the contour $\Gamma$ is the one depicted in Fig.~\ref{figure1}. We note that the poles on the integrand of~\eqref{Integral2} are located at $z=0$ and $z = \pm i m_{(i)j}$, with $j=1,2,\cdots,n_i$, and that complex quantities $m_{(i)j}$ always occur in conjugate pairs. Since there is only a finite number of poles, it is always possible to choose a contour $\Gamma$ which encompasses all the poles in the upper-half $\mathbb{C}$ plane.

The next step is to apply Cauchy's theorem to each pole inside $\Gamma$. Let us first suppose that the poles are all simple. Then,
\begin{eqnarray} 
I_i & = & - i\pi  +  2 i \pi  \sum_{j=1}^{n_i} \lim_{z \rightarrow i m_{(i)j}} \frac{m_{(i)j}^2 e^{izr}}{z (z + im_{(i)j})} \prod_{k \neq j} \frac{m_{(i)k}^2}{z^2 + m_{(i)k}^2} \nonumber   \\
& = & - i \pi \, \left( 1 + \sum_{j=1}^{n_i} e^{-r m_{(i)j}} \prod_{k\neq j} \frac{m_{(i)k}^2}{m_{(i)k}^2 - m_{(i)j}^2} \right) . \label{result_I_i}
\end{eqnarray}

Inserting~\eqref{result_I_i} into~\eqref{phi} we conclude that if there are only simple poles the non-relativistic potential is given by
\begin{eqnarray} \label{Potential_Simple}
\phi(r) & = & -GM \, \Bigg[ \frac{1}{r} - \frac{4}{3} \sum_{i=1}^{n_2} \frac{e^{-r m_{(2)i}}}{r} \prod_{j\neq i} \frac{m_{(2)j}^2}{m_{(2)j}^2 - m_{(2)i}^2}
\nonumber
\\
&& + \, \frac{1}{3} \sum_{i=1}^{n_0} \frac{e^{-r m_{(0)i}}}{r} \prod_{j\neq i} \frac{m_{(0)j}^2}{m_{(0)j}^2 - m_{(0)i}^2} \, \Bigg] \, .
\end{eqnarray}

Our derivation makes it clear that the expression~\eqref{Potential_Simple}, which was obtained for the first time in~\cite{Quandt-Schmidt} for the case $Q_2(k^2) \equiv 1$ and in~\cite{Newton-high} for non-trivial polynomials, also holds if the massive quantities $m_{(i)j}$ are complex numbers. It is now straightforward to verify that in such a scenario the potential $\phi(r)$ is still real. In fact, as a consequence of the fundamental theorem of algebra, the complex roots of $Q_i$ always occur in conjugate pairs. Let $m_{(i)j} = a_{(i)j} + i b_{(i)j}$ and $\overline{m}_{(i)j}$ be a pair of conjugate roots. Then,
%
%\begin{eqnarray}
%\text{i.} & & m_{(i)j}^2 \overline{m}_{(i)j}^2 = (a_{(i)j}^2 - b_{(i)j}^2)^2 + 4a_{(i)j}^2 b_{(i)j}^2;\\
%\text{ii.} &  &m_{(i)j}^2 - \overline{m}_{(i)j}^2 = 4 i a_{(i)j}^2 b_{(i)j}^2;\\
%\text{ii.} &  &(m_{(i)j}^2 - m_{(i)k}^2) (\overline{m}_{(i)j}^2 - m_{(i)k}^2) =  (a_{(i)j}^2 - b_{(i)j}^2 - m_{(i)k}^2)^2 + 4a_{(i)j}^2 b_{(i)j}^2.
%\end{eqnarray}
%
\begin{itemize}
\item[i.] $m_{(i)j}^2 \overline{m}_{(i)j}^2 = (a_{(i)j}^2 - b_{(i)j}^2)^2 + 4a_{(i)j}^2 b_{(i)j}^2$;
\item[ii.] $m_{(i)j}^2 - \overline{m}_{(i)j}^2 = 4 i a_{(i)j}^2 b_{(i)j}^2 $;
\item[iii.] $(m_{(i)j}^2 - m_{(i)k}^2) (\overline{m}_{(i)j}^2 - m_{(i)k}^2) =  (a_{(i)j}^2 - b_{(i)j}^2 - m_{(i)k}^2)^2 + 4a_{(i)j}^2 b_{(i)j}^2$.
\end{itemize}

Therefore, the terms in~\eqref{Potential_Simple} which involve exponentials of real quantities do themselves assume real values, since owed to the above relations i) and iii) both their numerator and denominator are real. On the other hand, the terms which contain exponentials of complex ``masses'' are true complex quantities. Notwithstanding, the sum of the term related to a complex $m_{(i)j}$ and the other associated to its conjugate turns out to be a real number. To prove this statement we define the notation $\overline{m}_{(i)j} \equiv m_{(i)j^\prime} \,$; then, using the relations ii) and iii) it follows that
\begin{eqnarray}
W_j & \equiv & e^{-r m_{(i)j}} \prod_{k\neq j} \frac{m_{(i)k}^2}{m_{(i)k}^2 - m_{(i)j}^2} + e^{-r {m}_{(i)j^\prime}} \prod_{k\neq j^\prime} \frac{m_{(i)k}^2}{m_{(i)k}^2 - {m}_{(i)j^\prime}^2} \nonumber \\
%& = & -\frac{m_{(i)j^\prime}^2}{4 i a_{(i)j}^2 b_{(i)j}^2} e^{-r m_{(i)j}}  \prod_{k\neq j,j^\prime} \frac{m_{(i)k}^2}{m_{(i)k}^2 - m_{(i)j}^2} \nonumber \\
%& & + \, \frac{m_{(i)j}^2}{4 i a_{(i)j}^2 b_{(i)j}^2} e^{-r m_{(i)j^\prime}} \prod_{k\neq j,j^\prime} \frac{m_{(i)k}^2}{m_{(i)k}^2 - {m}_{(i)j^\prime}^2} \nonumber \\
& = & A_j \exp(-r {m}_{(i)j})   + \overline{A_j} \exp(-r {\overline{m}}_{(i)j}) \nonumber \\
& = & 2 \text{Re} \left[ A_j \exp(-r {m}_{(i)j})\right]
%& = & 2 (\text{Re} \, A_j) \cos(r \, \text{Re} \, {m}_{(i)j}) + 2 (\text{Im} \, A_j) \sin (r \, \text{Im} \, {m}_{(i)j}),
\end{eqnarray}
where we have defined
\begin{equation*}
A_j \equiv \frac{i m_{(i)j^\prime}^2}{4 a_{(i)j}^2 b_{(i)j}^2} \prod_{k\neq j,j^\prime} \frac{m_{(i)k}^2}{m_{(i)k}^2 - m_{(i)j}^2}.
\end{equation*}
We conclude that the potential~\eqref{Potential_Simple} is always a real quantity, even if we allow complex poles in the propagator. Moreover, complex poles yield an oscillating behaviour to the potential, as noticed recently in~\cite{Modesto2016,Large} for particular cases of the polynomials $F_{i}^{(j)}$ in the Lagrangian~\eqref{Lagrangian}.

\section{Regularity of the potential at the origin}
\label{sec3}

Once proved that the potential is real even in the presence of simple complex poles, we shall prove its regularity at the origin. The behaviour of $\phi$ at small distances is dominated by the divergent factor $r^{-1}$, which appears in~\eqref{Potential_Simple} not only in the usual Newtonian term but also as the zero order term of the series expansion of the Yukawa-like potentials. Indeed, the singular terms which contribute to the potential occur in the factor
\begin{eqnarray} %%%%%%%%%%%%%%%%%%%%%%
\phi_\text{div}(r)= -\frac{GM}{r} \, \Bigg[ 1 - \frac{4}{3} \sum_{i=1}^{n_2} \prod_{j\neq i} \frac{m_{(2)j}^2}{m_{(2)j}^2 - m_{(2)i}^2} 
 +  \frac{1}{3} \sum_{i=1}^{n_0} \prod_{j\neq i} \frac{m_{(0)j}^2}{m_{(0)j}^2 - m_{(0)i}^2} \Bigg] \, .
\end{eqnarray}
If $\phi_\text{div}(r) \equiv 0$ the Newtonian singularity is cancelled by the massive modes.

Hence, in order to show that the potential is finite at the origin it suffices to verify that the relation~\cite{Quandt-Schmidt}
\begin{equation} \label{Relation}
\sum_{i} \prod_{j \neq i} \frac{a_j}{a_j - a_i} = 1
\end{equation}
is valid for any set $\lbrace a_i \rbrace_i$ of complex numbers. The following elementary identity is very useful in our task.

\newtheorem{thm}{Proposition} 
\begin{thm}
For any set of complex numbers $\lbrace a_i : i \in I \rbrace$, with $I = \lbrace 1, 2, \cdots, n \rbrace$, it holds that
\begin{equation} \label{eqproposicao}
\sum_{i} (-1)^{n+i} a_i^{n-1} \prod_{\substack{k<j \\ j,k \neq i}} (a_j - a_k) = {\prod_{i<j}} (a_j - a_i).
\end{equation}
\end{thm}

\newproof{pf}{Proof}
\begin{pf}
One should first recognize the right-hand side of~\eqref{eqproposicao} as the determinant of the $n$\textsuperscript{th}-order Vandermonde matrix~$\mathcal{V}_{n}$:
%This relation is trivial if one recognizes the right-hand side as the determinant of the $n$\textsuperscript{th}-order Vandermonde matrix~$\mathcal{V}_{n}$:
%
\begin{equation} \label{Vandermonde}
\det \mathcal{V}_{n} = \begin{vmatrix}
1      &  1     &  1     & \cdots & 1     \\
a_1    & a_2    & a_3    & \cdots & a_n   \\
a_1^2  & a_2^2  & a_3^2  & \cdots & a_n^2 \\
\vdots & \vdots & \vdots & \ddots & \vdots \\
a_1^{n-1}  & a_2^{n-1}  & a_3^{n-1}  & \cdots & a_n^{n-1} \\
\end{vmatrix} = {\prod_{i<j}} (a_j - a_i).
\end{equation}

Then, proceeding the cofactor expansion along its last raw,
\begin{eqnarray} \label{Vandermonde2}
\det \mathcal{V}_{n} & = & \sum_i (-1)^{n+i} a_i^{n-1} \det \mathcal{M}_{i} \\
& = & \sum_i (-1)^{n+i} a_i^{n-1} \prod_{\substack{k<j \\ j,k \neq i}} (a_j - a_k),
\end{eqnarray}
where the minors $\mathcal{M}_{i}$ related to the element $(n,i)$ are also Vandermonde matrices having their determinants calculated using the well-known relation~\eqref{Vandermonde}. This completes the proof.
\end{pf}

With this identity in hand, we carry out the proof of Formula~\eqref{Relation}. We start writing its left-hand side (LHS) as a ratio:
\begin{equation} \label{ratio}
\sum_{i=1}^n \prod_{j \neq i} \frac{a_j}{a_j - a_i} = \frac{\sum_i \prod_{j\neq i}  a_j (a_i - a_j) \prod_{k\neq i,j}  (a_j - a_k)}{\prod_{i\neq j} (a_i - a_j)} \equiv \frac{N}{D}.
\end{equation}
If we now apply the relation $a_j(a_i - a_j) = (a_i - a_j)(a_j - a_i) - a_i ( a_j - a_i)$ to each term $a_j(a_i - a_j)$ in the numerator $N$ it follows
\begin{eqnarray} \label{previous}
N & = & \sum_{i,m,C\lbrace k\rbrace} \Bigg[  (-a_i)^{n-m-1} \prod_{j = k_1}^{k_m} (a_i - a_j) (a_j - a_i)
  \nonumber \\
  & & \times \prod_{r \neq i,k_1,k_2,\cdots ,k_m}  (a_r - a_i) \prod_{\substack{s,t \neq i \\ s \neq t}} (a_s - a_t) \Bigg],
\end{eqnarray}
where the summation is carried out over all element indexes $i \in I = \lbrace 1, 2, \cdots n \rbrace$, every number $m= 0, 1, \cdots, n-1 \,$ and over every possible combination of $m$ indexes $k_1, k_2, \cdots, k_m~\in~I$; also, the notation implies the substitutions
\begin{itemize}
\item for $m=0$: $\,\, \prod_{j = k_1}^{k_m} (a_i - a_j) (a_j - a_i) \, \longmapsto \, 1$ ,
\item for $m=n-1$: $\,\, \prod_{r \neq i,k_1,k_2,\cdots ,k_m}  (a_r - a_i) \, \longmapsto \, 1$.
\end{itemize}
%
%\begin{eqnarray*}
%\text{for } & m=0, & \prod_{j = k_1}^{k_m} (a_i - a_j) (a_j - a_i) \mapsto 1 , \\
%\text{for } & m=n-1,  & \prod_{r \neq i,k_1,k_2,\cdots ,k_m}  (a_r - a_i) \mapsto 1  .
%\end{eqnarray*}

The expression~\eqref{previous} can be cast into a more useful form by defining the sets $K_m = \lbrace k_1, k_2, \cdots, k_m \rbrace$ if $m>0$  (while $K_0 = \emptyset$), $Y_{K_m} = \lbrace y_1, \cdots, y_{n-m} : y_\lambda \in I\setminus K_m \,, \text{ with } y_1<\cdots <y_{n-m}\rbrace$, $W_{\lambda,K_m} = Y_{K_m}\setminus \lbrace y_\lambda \rbrace$, 
and the ordering function
\begin{equation}
f(a_i,a_j) = \left\{ 
\begin{array}{l l}
a_i - a_j, & \quad \text{if $i < j$,}\\
a_j - a_i, & \quad \text{if $j < i$.}\\
\end{array} \right .
\end{equation}
Therefore,
\begin{eqnarray}
N & = & \sum_{m,C\lbrace k\rbrace} \sum_{\lambda = 1}^{n-m} \Bigg[ (-1)^{n-m+\lambda}  (-a_{y_\lambda})^{n-m-1} \prod_{j \in K_m} (a_{y_\lambda} - a_j) (a_j - a_{y_\lambda}) 
\nonumber \\
& & \times \prod_{r \in W_{\lambda,K_m}}  f(a_{y_\lambda},a_r) \prod_{\substack{s \neq t \\ s,t \neq y_\lambda}} (a_s - a_t) \Bigg] .
\end{eqnarray}

In the spirit of the Proposition~1, for each $m$ and each $K_m$ it holds that
\begin{equation}
\sum_{\lambda = 1}^{n-m} (-1)^{n-m+\lambda} (a_{y_\lambda})^{n-m-1} \prod_{t<s \in W_{\lambda,K_m}} (a_s - a_t) = \prod_{t<s \in Y_{K_m}} (a_s - a_t),
\end{equation}
because the LHS can be regarded as the determinant of a Vandermonde matrix of order $n-m$ whose elements are built from the set $\lbrace a_i : i \in I\setminus K_m \rbrace$. Thus, it is not difficult to verify that
\begin{eqnarray}
N & = & \sum_{m,C\lbrace k\rbrace}(-1)^{n-m-1}   {\prod_{s \neq t}} (a_s - a_t) \nonumber \\
& = & -{\prod_{s \neq t}} (a_s - a_t) \sum_{m=0}^{n-1} {n \choose m} (-1)^{n-m} \nonumber \\
& = & {\prod_{s \neq t}} (a_s - a_t) .
\end{eqnarray}

Substituting this result for $N$ into~\eqref{ratio} it follows the identity
\begin{equation} \label{proof}
\sum_i \prod_{j \neq i} \frac{a_j}{a_j - a_i} = \frac{\prod_{j\neq i} (a_i - a_j)}{\prod_{j\neq i} (a_i - a_j)} = 1.
\end{equation}
Hence, the divergent contribution of the potential $\phi_\text{div} (r)$ is null even if simple complex poles are taken into account.

\section{Potential with degenerate poles}
\label{sec4}

Up to this point we have only dealt with the situation of simple poles in the propagator. Nevertheless, the proof of the identity~\eqref{proof} shows that it remains valid if there are degenerate elements within the set $\lbrace a_i \rbrace_i$ --- that is, if it happens that $a_r = a_s$ for $r \neq s$. Indeed, the limit $a_r \longrightarrow a_s$ for any pair $(a_r, a_s)$ can be promptly evaluated with no dilemma, since the ``problematic'' term $a_r - a_s$ in the denominator of~\eqref{proof} is cancelled by the corresponding one which occurs in its numerator. This suggests that the potential is also finite in the case of higher-order poles, as we show in this section.

If there are poles of order greater than one in the propagator~\eqref{propagator}, the general procedure implemented in section~\ref{sec2} to calculate the potential remains valid, but now the residues at the multiple poles must be evaluated via the standard formula which requires $n-1$ derivations with respect to $z$ for a pole of order $n$.
%
%\begin{equation} \label{residue}
%\begin{split}
%\text{Res} & \left[ e^{izr} \prod_{t} \frac{m_{(i)t}^2}{z(z^2 + m_{(i)t}^2)} ; z = i m_{(i)j}\right]  = 
%\\& \lim_{z \rightarrow im_{(i)j}}  \frac{1}{(n-1)!} \frac{\partial^{n-1}}{\partial z^{N-1}} \left[ \frac{e^{izr}}{z (z + im_{(i)j})^n} \prod_{t \notin S} \frac{m_{(i)t}^2}{z^2 + m_{(i)t}^2} \right],
%\end{split}
%\end{equation}
%
%where $z = i m_{(i)j}$ is a pole of order $n$. Counting the number of poles from the order of the polynomial $Q_i(k^2)$, we may well consider as if we had a set $\lbrace m_{(i)s} : s \in S \rbrace$ of $N$ degenerate poles of the same ``mass'' $m_{(i)j}$ --- here $S$ is the set of indices corresponding to these poles. It is in this sense that the above expression must be understood.
%
The presence of higher derivatives in the evaluation of the residues makes the expression for the potential rather cumbersome. However, it always follows the structure defined by~\eqref{phi}:
\begin{equation} \label{structure}
\phi(r) = -\frac{GM}{r} \left[ 1 - \frac{4}{3} C_2(r) + \frac{1}{3} C_0(r) \right].
\end{equation}
For example, in the particular case $p=q=2$ (see Eq.~\eqref{Lagrangian}) %with one real simple massive mode and two degenerate ones in each sector, we have
%
%\begin{eqnarray}
%C_i(r) & = & \frac{m_{(i)2}^4 e^{-m_{(i)1}r}}{(m_{(i)1}^2 - m_{(i)2}^2)^2}  + e^{-m_{(i)2}r} \bigg(  \frac{m_{(i)1}^2 m_{(i)2} r}{2(m_{(i)1}^2 - m_{(i)2}^2)} 
%\nonumber \\ 
%& & - \, \frac{m_{(i)1}^2 m_{(i)2}^2}{(m_{(i)1}^2 - m_{(i)2})^2} +  \frac{m_{(i)1}^2}{m_{(i)1}^2 - m_{(i)2}^2} \bigg)   .
%\end{eqnarray}
%
with all the three modes in each sector having the same mass $m_i$,
\begin{equation}
C_i(r) = e^{-m_i r} + \frac{5}{8} m_i r e^{-m_i r} + \frac{1}{8} m_i^2 r^2 e^{-m_i r} .
\end{equation}

The same expression can be obtained starting from the simple pole formula~\eqref{Potential_Simple} by taking the limit
\begin{equation}
C_i(r) = \lim_{m_{(i)3}, m_{(i)2} \rightarrow m_{(i)1}} \left( \sum_{j=1}^{3} e^{-r m_{(i)j}} \prod_{k\neq j} \frac{m_{(i)k}^2}{m_{(i)k}^2 - m_{(i)j}^2} \right).
\end{equation}
%
%in the case of degeneracy two, and the limit $m_{(i)3}, m_{(i)2} \longrightarrow m_{(i)1}$ for the scenario with a third-order pole.
%
%\begin{equation}
%C_i(r) = \lim_{m_{(i)3}, m_{(i)2} \rightarrow m_{(i)1}}  \left( \sum_{j=1}^{3} e^{-r m_{(i)j}} \prod_{k\neq j} \frac{m_{(i)k}^2}{m_{(i)k}^2 - m_{(i)j}^2} \right) ,
%\end{equation}
%
%for the scenario of three degenerate poles.

More generally, it is possible to compute the potential in the presence of degenerate poles simply by taking the corresponding limit in the usual expression~\eqref{Potential_Simple} for simple poles, instead of evaluating the residues at multiple poles. Even though calculating the limit may be a more tedious procedure, it is useful for proving that the regularity of the potential is preserved.

In fact, let us suppose that we have a model with $n_i$ spin-$i$ massive modes, among which $s_i$ are degenerate --- that is $m_{(i)j} = m_i$ if and only if $j \leq s_i$. Treating each of the degenerate poles as a simple one, infinitesimally close to $m_i$, we have the spin-$i$ contribution to the potential proportional to
\begin{eqnarray}
C_i(r) & = & \lim_{\varepsilon_j \rightarrow 0} \Bigg[  \sum_{j=1}^{s_i} e^{-r (m_i + \varepsilon_j)} \prod_{k\neq j} \frac{m_{(i)k}^2}{m_{(i)k}^2 - (m_i + \varepsilon_j)^2} 
\nonumber \\
& & + \, \sum_{j=s_i+1}^{n_i} e^{-r m_{(i)j}} \prod_{k\neq j} \frac{m_{(i)k}^2}{m_{(i)k}^2 - m_{(i)j}^2} \Bigg] .
\end{eqnarray}
At small distances, the only divergent contribution\footnote{The divergences related to the limit $\varepsilon_j \longrightarrow 0$ keeping $r$ constant are all cancelled, being the potential well-defined, as it can be deduced from the standard expression for the residue at a multiple pole.} comes from the terms proportional to $r^{-1}$, related to the zero order approximation of the Yukawa-like potentials. % (First and higher order terms of the expansion of the exponentials yield finite contributions to the potential of the order $r^0$ or higher.)
Thus, sufficiently close to the origin it follows that
%
%\begin{equation}
%C_i(r)  \longrightarrow  \lim_{m_{(i)1},\cdots,m_{(i)s_i} \rightarrow m_{i}} \sum_{j=1}^{n_i} \prod_{k\neq j} \frac{m_{(i)k}^2}{m_{(i)k}^2 - m_{(i)j}^2} = 1,
%\end{equation}
\begin{equation}
C_i(r)  \longrightarrow  \lim_{\substack{ m_{(i)t} \rightarrow m_{i} \\ \forall t \leq s_i}} \sum_{j=1}^{n_i} \prod_{k\neq j} \frac{m_{(i)k}^2}{m_{(i)k}^2 - m_{(i)j}^2} + \mathcal{O}(r) = 1 + \mathcal{O}(r),
\end{equation}
where we have used~\eqref{proof}. This result holds, of course, independently of the number of multiple poles and their real or complex nature. Also, since the potential for simple poles given by~\eqref{Potential_Simple} is always a real quantity, it remains real when the corresponding limit is taken to allow for complex degenerate roots of~$Q_i$.

\section{More on the cancellation of the Newtonian singularity}
\label{sec5}

In view of the results obtained in the previous sections it is worth addressing some words on the mechanism behind the cancellation of the singularity of the potential. With respect to Eq.~\eqref{structure}, we have seen that each quantity $C_i(r)$ depends only on the polynomial $Q_i$ defined by~\eqref{Polynomials}; and at short distances it tends to $1+\mathcal{O}(r)$ independently of the number of spin-$i$ modes. As a consequence, the potential remains finite if the degrees of $Q_0$ and $Q_2$ are not the same, that is if $p \neq q$ in the context of Eq.~\eqref{Lagrangian}. This suggests that the absence of Newtonian singularities in the classical gravitational potential occurs in the more general class of theories than the ones which are (super)renormalizable, contrary to the conjecture formulated in~\cite{Newton-high,Accioly13_riddle}.

As an example, consider the scenario in which $p = 1$ and $q=0$ described by the Lagrangian~\eqref{Lagrangian} with the polynomials $F_1 = \frac{a_0}{2} + \frac{a_1}{2} \square$ and $F_2 = \frac{b_0}{2}$. 
For the sake of simplicity, let us suppose that $a_0$, $a_1$ and $b_0$ are such that the massive poles are all real. Then, the masses of the extra modes are given by
\begin{equation*}
m_{0\pm}^2 = \frac{3a_0 + b_0 \pm \sqrt{(3a_0 + b_0)^2 - 24 a_1 \kappa^{-2}}}{6a_1}, \quad m_{2}^2 = \frac{4}{|b_0|\kappa^2}.
\end{equation*}
In this scenario, the potential~\eqref{Potential_Simple} boils down to
\begin{equation*}
\phi(r) = - \frac{GM}{r} \left[   1 - \frac{4}{3} e^{-m_{2} r} +  \frac{1}{3} \left(  \frac{m_{0-}^2e^{-m_{0+} r}}{m_{0-}^2 - m_{0+}^2} 
 +  \frac{m_{0+}^2e^{-m_{0-} r}}{m_{0+}^2 - m_{0-}^2}  \right) \right] ,
\end{equation*}
which is finite at the origin, as it can be easily verified.
%A moment's reflection reveals that the potential is finite, in fact
%\begin{equation}
%\phi(r) \xrightarrow{r \rightarrow 0} GM \left[ - \frac{4}{3} m_{(2)} + \frac{1}{3} \frac{m_{(0)+}m_{(0)-} \left( m_{(0)+} - m_{(0)-}\right)  }{m_{(0)+}^2 - m_{(0)-}^2}\right] .
%\end{equation}

In this example, $m_{2}$ and $m_{0+}$ are ghost excitations, while $m_{0-}$ is a healthy mode~\cite{Newton-high}. So, the singularity is cancelled despite the fact that there is no healthy massive tensor mode to balance the repulsive force of the scalar ghost of mass $m_{0+}$.

This observation gives a negative answer to the conjecture proposed in~\cite{Newton-high} which stated that the mechanism behind the cancellation of the singularity is the matching of the number of ghost and healthy modes between the spin-2 and spin-0 massive sectors. The authors argued that only if $p = q$ the potential would be regular, because the number of massive excitations in each sector would be the same and to each ghost mode in the scalar sector there would be a non-ghost tensor one, and vice versa. Since in gravity ghosts are associated to a repulsive interaction~\cite{Newton-high}, this balance between their repelling forces and the attractive ones (owed to healthy modes) is what would regularize the potential. This principle is correct, however in a model like~\eqref{N_OrderGravity} it is not the \emph{number} of ghost and non-ghost massive excitations which guarantee the regularity. Instead, it is the presence of \emph{at least one} massive mode in each sector.

We know that in the case the propagator has only real poles, the lightest massive tensor one is a ghost mode, while the lightest scalar is a non-ghost~\cite{Newton-high}. Also, because to each massive tensor excitation there is a scalar one, in a gravity model such as~\eqref{N_OrderGravity} the simplest form of having a regular potential is to have one spin-2 and one spin-0 massive mode. With the further requirement  that the spectrum does not have tachyons, this corresponds precisely to Stelle's renormalizable model~\cite{Stelle77}. The insertion of sixth- or higher-derivative terms which are quadratic in curvature allows more exotic configurations with complex and/or degenerate poles, but no new dynamically independent field. No matter the number of excitations each sector has, their dynamics are coupled in such a manner that the resulting contribution $C_i$ at small distances always tend to the unit (see~\cite{Large} for a specific discussion on the sixth-order gravity). In fact, the relevant terms on the propagator~\eqref{propagator} to the determination of the weak field generated by a static point-like mass are those proportional to $P^{(2)}$ and $P^{(0-s)}$. Then, if one expands, e.g., the coefficients of the spin-2 projector in partial fractions it yields
%
%In order to close this work we comment briefly some possibilities of avoiding the cancellation of the Newtonian singularity, namely:
%
%\begin{itemize}
%\item[i.] changing the coefficients $-4/3$ and $+1/3$ of Eq.~\eqref{structure} that add to $-1$;
%\item[ii.] changing the limit $C_{0,2}(r) \longrightarrow 1$.
%\end{itemize}

%In what concerns the first possibility, we recall that the coefficients $4/3$ and $1/3$ are related to the fact that we consider a 4-dimensional space-time (see for example Ref.~\cite{Accioly15}), which suggests that the regularity of the potential is related to the dimension, like renormalizability. On the other hand, the second possibility could be achieved if the dynamics of the massive modes could be decoupled. It seems, however, that the insertion of sixth- or higher-derivative terms which are quadratic in curvature only yields higher-order equations of motions, but no new dynamically independent field.
%In order to prove this statement we return to the propagator~\eqref{propagator} and notice that the relevant terms to the determination of the weak field generated by a point-like mass in rest are those proportional to the operators $P^{(2)}$ and $P^{(0-s)}$. Then, if one expands, for example, the coefficients of the spin-2 projector in partial fractions it yields
%
\begin{equation}
\frac{1}{k^2 Q_2(k^2)} = \frac{1}{k^2} + \sum_i \frac{c_i}{k^2 - m_{(2)i}^2} ,
% \quad c_i = - \prod_{j \neq i} \frac{m_{(2)j}^2}{m_{(2)j}^2 - m_{(2)i}^2}.
\end{equation}
where the coefficients $c_i$ depend on all the quantities $m_{(2)j}$. Therefore, we cannot regard the spin-2 excitation of mass $m_{(2)i}$ as a totally independent field, since in its equations of motion there is a factor which depends on the masses of the other tensor modes. In other words, the amount each of these fields will contribute to the potential at the origin is not an arbitrary quantity to be determined by the field itself, but vary according to how the other fields of the same spin contribute. The only field which is totally decoupled corresponds to the graviton.

\section{Conclusions}
\label{conclusions}

In the present work we have proved that the classical potential associated to the higher-derivative gravity model described by the action~\eqref{N_OrderGravity} is a real quantity even if we allow for complex and/or multiple massive poles in the propagator. Moreover, it was shown that the expression for the potential has the general form given by the Eq.~\eqref{structure} %
%
%\begin{equation}  \label{conclusion}
%\phi(r) = -\frac{GM}{r} \left[ 1 - \frac{4}{3} C_2(r) + \frac{1}{3} C_0(r) \right]
%\end{equation}
%
with $C_{0,2}(r) \longrightarrow 1+\mathcal{O}(r)$ at short distances, making the potential regular at the origin regardless of the complex or real nature of the poles, their degeneracies or the number of massive excitations for each spin. This result is particularly interesting since it applies to the superrenormalizable and unitary versions of quantum gravity~\cite{ModestoShapiro16,Modesto2016}.

Also, it was shown that the cancellation of the Newtonian singularity in the model~\eqref{N_OrderGravity} is a consequence of having all the possible curvature-squared terms in the action, without making reference to the order of the polynomials $F_i(\square)$. Adding more derivatives introduces more degrees of freedom, but the dynamics of the massive modes are coupled in such a manner that the regularity of the potential is preserved, even in non-renormalizable models. Finally, the absence of singularity in the potential of the non-local gravity~\cite{Tseytlin95Tomboulis} may be viewed as the limit scenario of our result on the polynomial action.

\section*{Acknowledgements}
The author is very grateful to A. Accioly and I.L. Shapiro for the fruitful discussions; and to CNPq, for supporting his Ph.D. project.

%\section*{References}





\end{document}